\documentclass[12pt]{article}

\usepackage{amsmath}
\usepackage[doublespacing]{setspace}
\usepackage{graphicx}
\usepackage{pslatex}
\usepackage[margin = 1in]{geometry}
\usepackage{color}
\usepackage{caption}
\usepackage{subcaption}
\usepackage{booktabs}
\usepackage[colorlinks = true, urlcolor=black, citecolor=black,linkcolor=black]{hyperref}
\usepackage{enumitem}
\usepackage{wrapfig}
\usepackage{authblk}
\usepackage{siunitx}
\usepackage[superscript]{cite}
\usepackage{multicol}
\usepackage{multirow}
\usepackage{booktabs}
\usepackage[toc, page]{appendix}
\usepackage{pdflscape}
\usepackage{afterpage}
\usepackage{sectsty}
\sectionfont{\fontsize{12}{12}\selectfont}
\subsectionfont{\fontsize{10}{10}\selectfont}

\newcommand{\PM}{$\text{PM}_{2.5}$}
\newcommand{\SO}{$\text{SO}_{2}$}

\begin{document}

\title{A Source-Oriented Approach to Coal Power Plant Emissions Health Effects}

\author[1]{Kevin Cummiskey\thanks{kevin.cummiskey@westpoint.edu}}
\author[4]{Chanmin Kim}
\author[2]{Christine Choirat}
\author[2]{Lucas R.F. Henneman}
\author[3]{Joel Schwartz}
\author[2]{Corwin Zigler}


\affil[1]{Department of Mathematical Sciences, West Point, NY 10996}
\affil[2]{Department of Biostatistics, Harvard T.H. Chan School of Public Health, 677 Huntington Ave, Boston, MA 02115}
\affil[3]{Department of Environmental Health, Harvard T.H. Chan School of Public Health, Boston, MA}
\affil[4]{Department of Biostatistics, Boston University School of Public Health, Boston, MA}

\begin{singlespacing}
\maketitle
\end{singlespacing}

\section*{Summary}
\subsection*{Background}
There is increasing focus on whether air pollution originating from different sources has different health implications. In particular, recent evidence suggests that fine particulate matter (\PM{}) with chemical tracers suggesting coal combustion origins is especially harmful. Augmenting this knowledge with estimates from causal inference methods to identify the health impacts of \PM{} derived from specific point sources of coal combustion would be an important step towards informing specific, targeted interventions. 

\subsection*{Methods}
We investigated the effect of high-exposure to coal combustion emissions from 783 coal-fired power generating units on ischemic heart disease (IHD) hospitalizations in over 19 million Medicare beneficiaries residing at 21,351 ZIP codes in the eastern United States.  We used InMAP, a newly-developed, reduced-complexity air quality model to classify each ZIP code as either a high-exposed or control location.  Our health outcomes analysis uses a causal inference method - propensity score matching - to adjust for potential confounders of the relationship between exposure and IHD.  We fit separate Poisson regression models to the matched data in each geographic region to estimate the incidence rate ratio for IHD comparing high-exposed to control locations.       

\subsection*{Findings}  
High exposure to coal power plant emissions and IHD were positively associated in the Northeast $(IRR = 1.08, 95\%\text{ CI} = 1.06, 1.09)$ and the Southeast ($IRR$ = 1.06, 95\% CI = $1.04, 1.08$).  No significant association was found in the Industrial Midwest $(IRR = 1.02, 95\% \text{ CI} = 1.00, 1.04)$, likely the result of small exposure contrasts between high-exposed and control ZIP codes in that region.   

\subsection*{Interpretation}
This study provides targeted evidence of the association between emissions from specific coal power plants and IHD hospitalizations among Medicare beneficiaries.

\subsection*{Funding} 
USEPA grant RD-835872-01, NIH grant R01ES026217, and HEI grant 4953.


\section{Introduction}

Long-term exposure to ambient fine particulate matter of less than 2.5 micrometers in diameter (\PM{}) has been associated with increased mortality and morbidity related to various cardiovascular and respiratory conditions \cite{lepeule2012chronic, dockery1993association, pope2002lung, pope1995particulate, pope2004cardiovascular}.   While the association between overall \PM{} exposure and health outcomes is well-documented, \PM{} varies in chemical composition and there is increasing evidence particles originating from different sources impact health differently \cite{laden2000association, ostro2010long, lippmann2014toxicological, thurston2016ischemic}. The National Academy of Sciences identified this area of research as a top priority because it ``could result in targeted control strategies that would specifically address these sources having the most significant effects on public health"\cite{national2004research}.  

Most existing research on the comparative health effects of different sources of \PM{} relies on source apportionment methods to apportion \PM{} mass measured at population locations into broad source categories, based in part on the availability of certain chemical tracers associated with specific types of sources. 
Despite the value of using source apportionment in health effects studies, such methods are limited by their reliance on broad source categories identified by chemical tracers.  Rather, it may be valuable to estimate the health impact associated with exposure to a specific set of point sources, for example, all coal-fired power plants operating in the United States. The ability to evaluate health impacts associated with specific point sources is an important step towards providing policy-relevant information that can inform specific, targeted interventions.  Moreover, observational epidemiological studies are particularly challenged by the threat of confounding; areas exhibiting high exposure may share important differences from areas with low exposure.  The ubiquity of potential confounding in studies of air pollution warrants the use of statistical methods anchored to causal inference methodology designed to more explicitly address the threat of confounding. 

The goal of this paper is to refine and complement the existing evidence of the health burden of coal combustion with a novel analysis approach that uses a reduced complexity air quality model to characterize exposure and combines this with statistical methods for causal inference and confounding adjustment to isolate the association between public health and emissions from a specific set of coal-fired power plants. Specifically, we estimate the association between a United States ZIP code having high exposure to coal combustion emissions and ischemic heart disease (IHD), mortality from which has been previously linked to long-term exposure to \PM{} \cite{miller2007long, hoek2013long, thurston2016ischemic}. Our study of IHD events offers additional granularity to the study of long-term \PM{} exposure and IHD which has, to date, focused on mortality.


We characterize coal power plant \PM{} exposure with a recently-developed, reduced-complexity air quality model called the Intervention Model for Air Pollution (InMAP, available at \url{http://spatialmodel.com/inmap/}).\cite{tessum2017inmap}  Specifically, we use InMAP to quantify the influence of coal emissions from 783 coal-fired power generating units at over 21,000 ZIP code locations in the Northeast, Industrial Midwest, and Southeast United States.  While InMAP is used to characterize coal emissions exposure, ZIP code level measures of total ambient \PM{} mass concentrations come from state-of-the-art predictions from Di et al (2016), derived from a neural-network combining information from monitoring data, land-use regression, remote-sensing satellite data, and GEOS-Chem simulations \cite{di2016assessing}. We link both measures with health outcomes available among 19,726,981 Medicare beneficiaries living in these regions to isolate the association between exposure to coal power plants and IHD events.  

A key component of our analysis approach is the use of propensity score matching, a method for confounding adjustment with advantages over traditional regression models. Grounding the statistical analysis to an explicit causal inference method such as propensity score matching is essential for informing policy interventions, particularly in observational epidemiological studies of air pollution \cite{dominici2017best}. 
The matching procedure is specifically designed to identify and mitigate the threat of confounding by framing the investigation as a hypothetical controlled experiment where each ZIP code is regarded as either high-exposed to coal power plant emissions or a control location.

To our knowledge, the combination of a reduced-complexity air quality model with modern statistical methods for causal inference and confounding adjustment deployed in an analysis of Medicare health outcomes and state-of-the-art data fused estimates of \PM{} represents the largest scale study to date of the health impacts of coal power plant pollution emissions.

\section{Methods}
\subsection{Data Sources and Study Population}
We compiled basic background and demographic information from the Center for Medicare and Medicaid Services on 19,726,981 Medicare beneficiaries residing in 21,351 ZIP codes in the Northeast, Industrial Midwest, and Southeast regions of the United States in 2005.  These three regions account for most coal power generation in the United States and have been subjected to national regulations intended to limit interstate transfer of air pollution emissions.  An IHD event was defined as a hospital admission with a primary discharge diagnosis of ICD-9 410-414, or 429. Population demographic data was augmented with information from the US Census Bureau (year 2000) and the CDC Behavioral Risk Factor Surveillance System.  Total annual emissions data for 783 coal-fired generating units was obtained from continuous emissions monitors provided in the EPA's Air Markets Program Data (AMPD).  Power plant stack features (height, diameter) were obtained from the EPA's National Emissions Inventory (NEI 2014).  

For a secondary analysis, we obtained \PM{} exposure predictions for 2005, from Di et al (2016), derived from a neural-network combining information from monitoring data, land-use regression, remote-sensing satellite data, and GEOS-Chem simulations \cite{di2016assessing}. 
Ultimately, the analysis data set contained data on 21,351 U.S. ZIP codes, each having a measure of IHD hospitalization rate in 2005, measures of population demographics, weather, and total \PM{} concentration, and a measure of coal power plant emissions exposure. 


\subsection{Classifying Coal Power Plant Exposure Using InMAP}
To create the primary exposure metric, we classified each ZIP code as either a high-exposed location to coal power plant emissions, or a control location, by combining data on power plant emissions with the results from InMAP.  InMAP uses output from a widely-used chemical transport model, WRF-Chem, to estimate changes in annual average \PM{} concentrations on a variable spatial grid attributable to annual changes in precursor emissions at user-prescribed locations. 

Total annual \SO{} emissions during 2005, geographic coordinates, and stack features for each coal-fired generating unit were input into InMAP.  For each location on the variable spatial grid, InMAP estimated the total annual change in \PM{} attributable to a 100\% emissions reduction for all coal generating units. These grid estimates were aggregated to obtain ZIP code estimates. 
While the InMAP output can be interpreted as the annual change in $\SI{}{\micro\gram}/\text{m}^3$ of \PM{} concentration, we refer to these estimates as the influence of coal emissions and use these values simply to classify ZIP codes (recall that total \PM{} mass predictions come from \cite{di2016assessing}). Based upon the distribution of these influences, an appropriate cutoff to classify locations as either high-exposed or control was selected, with this cutoff varied in sensitivity analyses.  

\subsection{Confounding Adjustment with Propensity Score Matching}
Isolating the association between high exposure to coal power plant emissions and IHD hospitalizations by comparing high-exposed to control locations requires adjusting for various population and climatological factors, also referred to as confounders, that differ between locations in both exposure groups.  

One common tool for confounding adjustment in such settings is propensity score matching, which, in this context, attempts to match high-exposed locations to control locations that are comparable on the basis of possible confounding factors.  Towards this goal, we estimated propensity scores for each ZIP code as the predicted probability of being high-exposed from a logistic regression model that included a broad set of covariates including socioeconomic and demographic variables, smoking rates, weather attributes, and characteristics of the Medicare population.  See Appendix \ref{app:data} for specific covariates included in the propensity score model.  
A secondary analysis augments propensity scores to include overall \PM{} mass.  

After propensity score estimation, each high-exposed location was matched to a control in the same region and with similar estimated propensity score (see Appendix \ref{app:matching} for details). After matching, the threat of confounding can be empirically assessed by checking whether covariate distributions are comparable in matched locations, before conducting any analysis of IHD hospitalizations.

\subsection{Model for IHD Hospitalizations}
The outcome of interest is the number of IHD hospitalizations in the Medicare population in 2005 in each ZIP code.  Poisson regression was used to estimate the incident rate ratio (IRR) for IHD hospitalization comparing high-exposed to control locations in the propensity score matched data set.  All models included the covariates from the propensity score model in the IHD model to adjust for residual confounding remaining after the matching process. To reflect the regional nature of air pollution, a separate Poisson regression model was fit to the matches for each region. 

\subsection{Sensitivity Analyses}
We present three sensitivity analyses. To assess the sensitivity to the exposure cutoff defining high-exposed ZIP codes, we performed the analysis under a range of different cutoffs delineating high coal emissions exposed ZIP codes from controls.  To assess sensitivity to the specific propensity score method for confounding adjustment, we analyzed the data by stratifying (instead of matching) locations on propensity score quintile and included an indicator of propensity score quintile in the model for IHD.  
This method typically results in many more locations in the IHD analysis and helps assess sensitivity of results to the specific locations chosen in the matching process.

Finally, to more directly address the prospect of unmeasured spatial confounding that could arise if there are unmeasured differences between ZIP codes that are located far from one another, an alternative method, Distance Adjusted Propensity Score Matching (DAPSm), was used to construct a matched data set based on estimated propensity scores and geographic proximity of matches (see Appendix \ref{app:DAPS}). \cite{doi:10.1093/biostatistics/kxx074} 


\subsection{Secondary Analysis: Adjusting for \PM{}}
Importantly, results of the primary analysis do not adjust for total \PM{} mass in the propensity score or outcome models.  As a consequence, estimates of the association between high coal emissions exposure and IHD include both the impact of elevated coal emissions themselves and any associated increase in overall ambient \PM{} mass.

As a secondary analysis, we augment the propensity score model to include the annual average ambient \PM{} concentration at each ZIP code. This analysis estimates the increased association between coal emissions exposure and IHD among ZIP codes that have been matched to have similar annual average total \PM{} mass.  Without restrictive assumptions, this analysis is difficult to interpret, as it amounts to adjusting for a variable ``on the causal pathway'' between coal emissions influence and IHD hospitalization; some areas may have total \PM{} mass that is largely a consequence of the coal influence exposure.  Under the assumption that annual average total \PM{} mass is not measurably affected by the high coal exposure metric (after adjusting for other variables in the propensity score), estimates in this secondary analysis address whether elevated coal power plant emissions are associated with IHD among areas with the same annual average total \PM{} mass.  Such a result would indicate differential health impact of coal-derived \PM{} relative to other sources that might make up total \PM{} mass; even for ZIP codes with the same annual ambient \PM{} concentration, those with high coal emissions exposure exhibit different health impacts. Appropriateness of this restrictive assumption relates to whether annual average total \PM{} mass, which is generally a consequence of many sources (including, but not limited to power plants), is not measurably affected by the high coal exposure metric, after adjusting for other variables in the propensity score.  Such may be the case, for example, in areas where \PM{} derived from coal power plant emissions represents a small proportion of overall ambient average \PM{} mass relative to other local pollution sources, or where coal emissions exposure determines the chemical composition of the ambient average but not the total mass. 

\subsection{Role of the funding source}

The funders had no role in the study design, data collection, analysis, interpretation, or writing of the report. The corresponding author had access to all study data and final responsibility for the decision to submit the report for publication.

\section{Results}

The study population experienced 537,369 IHD events in over 18 million person-years, a rate of 2856 events per 100,000 person-years.  These rates were 2657 per 100,000 in the Northeast, 3025 per 100,000 in the Industrial Midwest, and 2922 per 100,000 in the Southeast.  Table \ref{tab:rawdata} contains descriptive statistics of all covariates. 

\afterpage{%
\newgeometry{left = 1in, right = 0.25in, top = 0.25in, bottom = 0.25in}
\begin{landscape}

\begin{minipage}{10.5in}
\makebox[\textwidth][c]{
\renewcommand{\arraystretch}{0.75}
\begin{tabular}{@{} lllllll @{}}
  \toprule
 & \multicolumn{2}{c}{Industrial Midwest} & \multicolumn{2}{c}{Northeast} & \multicolumn{2}{c}{Southeast} \\ 
  \midrule
 &  Controls & High-exposed & Controls & High-exposed & Controls & High-exposed \\\midrule
Number of ZIP codes & 5576 & 1347 & 3852 & 3861 & 5298 & 1417 \\ 
  IHD events  & 25.44 (39.34) & 22.03 (34.63) & 19.62 (28.53) & 27.88 (44.21) & 27.16 (38.59) & 27.35 (31.85) \\ 
  Person-years  & 845.66 (1258.85) & 707.94 (1086.57) & 795.01 (1102.47) & 992.81 (1481.91) & 912.10 (1267.69) & 999.36 (1132.52) \\ 
  \PM{}  & 13.96 (2.06) & 15.10 (1.34) & 11.34 (1.79) & 13.98 (1.45) & 12.97 (2.20) & 14.53 (1.57) \\ 
  logPop  & 8.07 (1.66) & 8.04 (1.62) & 8.27 (1.49) & 8.44 (1.73) & 8.63 (1.53) & 9.00 (1.40) \\ 
  PctUrban  & 0.36 (0.42) & 0.33 (0.40) & 0.44 (0.44) & 0.55 (0.44) & 0.43 (0.43) & 0.47 (0.42) \\ 
  PctBlack  & 0.05 (0.13) & 0.04 (0.11) & 0.04 (0.10) & 0.12 (0.19) & 0.22 (0.24) & 0.22 (0.23) \\ 
  PctHisp  & 0.02 (0.06) & 0.01 (0.02) & 0.04 (0.08) & 0.04 (0.08) & 0.05 (0.10) & 0.03 (0.05) \\ 
  PctHighSchool  & 0.39 (0.10) & 0.41 (0.10) & 0.33 (0.10) & 0.35 (0.13) & 0.33 (0.08) & 0.30 (0.09) \\ 
  MedianHHInc  & 40.19 (14.85) & 36.42 (11.51) & 46.83 (19.34) & 47.92 (20.72) & 33.64 (11.64) & 40.15 (13.16) \\ 
  PctPoor  & 0.12 (0.10) & 0.13 (0.09) & 0.10 (0.08) & 0.10 (0.08) & 0.17 (0.10) & 0.13 (0.08) \\ 
  PctFemale  & 0.50 (0.04) & 0.51 (0.04) & 0.51 (0.03) & 0.51 (0.04) & 0.51 (0.04) & 0.51 (0.04) \\ 
  PctOccupied  & 0.88 (0.13) & 0.89 (0.10) & 0.84 (0.17) & 0.90 (0.11) & 0.87 (0.10) & 0.89 (0.09) \\ 
  PctMovedIn5  & 0.40 (0.11) & 0.41 (0.11) & 0.39 (0.10) & 0.39 (0.13) & 0.46 (0.11) & 0.47 (0.12) \\ 
  MedianHValue  & 92.34 (55.44) & 81.91 (33.19) & 137.63 (103.43) & 138.67 (109.63) & 85.60 (54.02) & 106.03 (55.61) \\ 
  smokerate2000  & 0.28 (0.03) & 0.30 (0.03) & 0.25 (0.03) & 0.25 (0.04) & 0.28 (0.03) & 0.26 (0.03) \\ 
  avtmpf  & 283.58 (2.04) & 285.40 (1.08) & 281.61 (1.72) & 284.80 (1.69) & 292.46 (3.45) & 289.34 (1.68) \\ 
  avrelh  & 0.01 (0.00) & 0.01 (0.00) & 0.01 (0.00) & 0.01 (0.00) & 0.01 (0.00) & 0.01 (0.00) \\ 
  mean\_age  & 74.91 (1.31) & 74.54 (1.33) & 75.15 (1.27) & 75.21 (1.48) & 74.72 (1.29) & 74.59 (1.13) \\ 
  Female\_rate  & 0.56 (0.05) & 0.55 (0.06) & 0.55 (0.06) & 0.56 (0.06) & 0.56 (0.05) & 0.58 (0.05) \\ 
  White\_rate  & 0.95 (0.14) & 0.96 (0.11) & 0.94 (0.13) & 0.86 (0.20) & 0.80 (0.21) & 0.80 (0.21) \\ 
   \hline
\end{tabular}
}
\captionof{table}{Mean (standard deviation) of ZIP code-level variables in the raw dataset. See Table \ref{tab:covariates} in the Appendix for variable definitions.}
\label{tab:rawdata}
\end{minipage}
\end{landscape}
\clearpage
\aftergroup\restoregeometry%
}

\subsection{Exposure Classification with InMAP}

\begin{figure}
\centering
\begin{subfigure}{0.58\textwidth}
\includegraphics[trim = {2.1in 3.45in 0cm 4.25in}, clip]{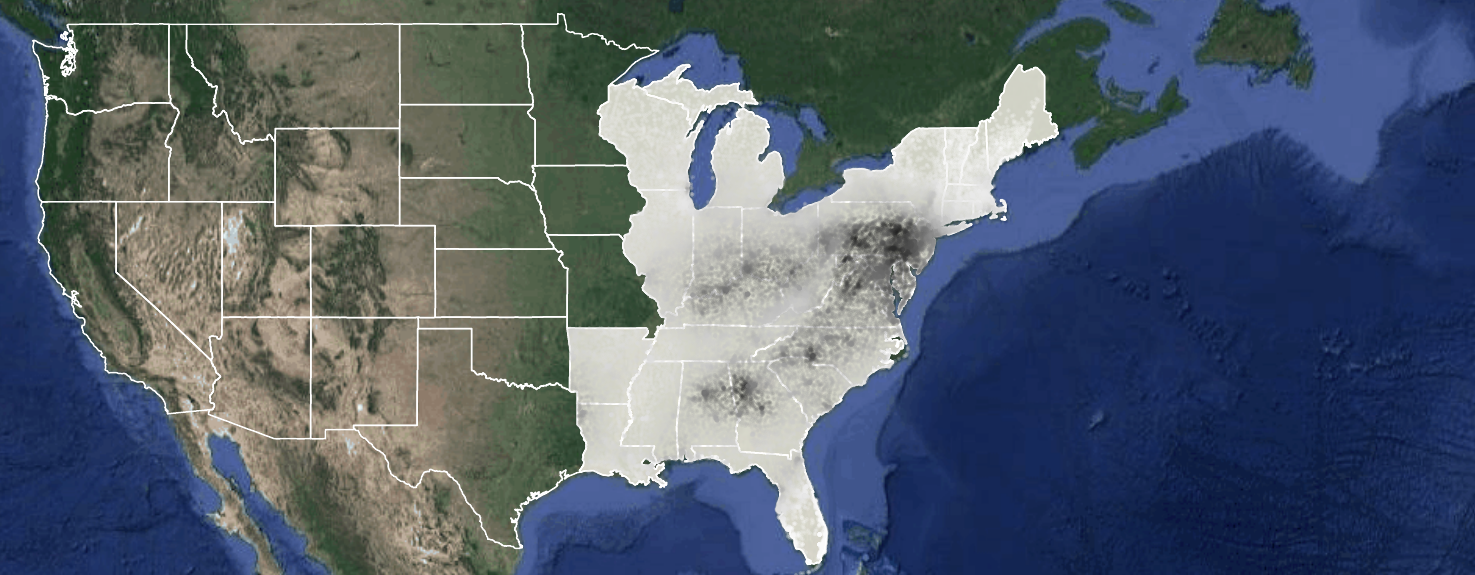}
\end{subfigure}%
~
\begin{subfigure}{0.36\textwidth}
\vspace{0.37in}
\includegraphics{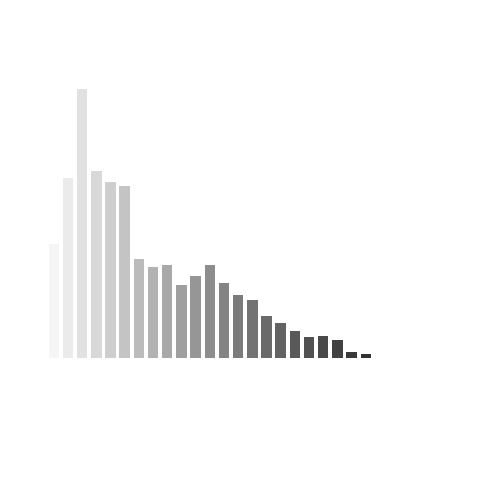}
\end{subfigure}%
\hfill
\begin{subfigure}{0.58\textwidth}
\includegraphics[trim = {2.1in 3.45in 0cm 4.25in}, clip]{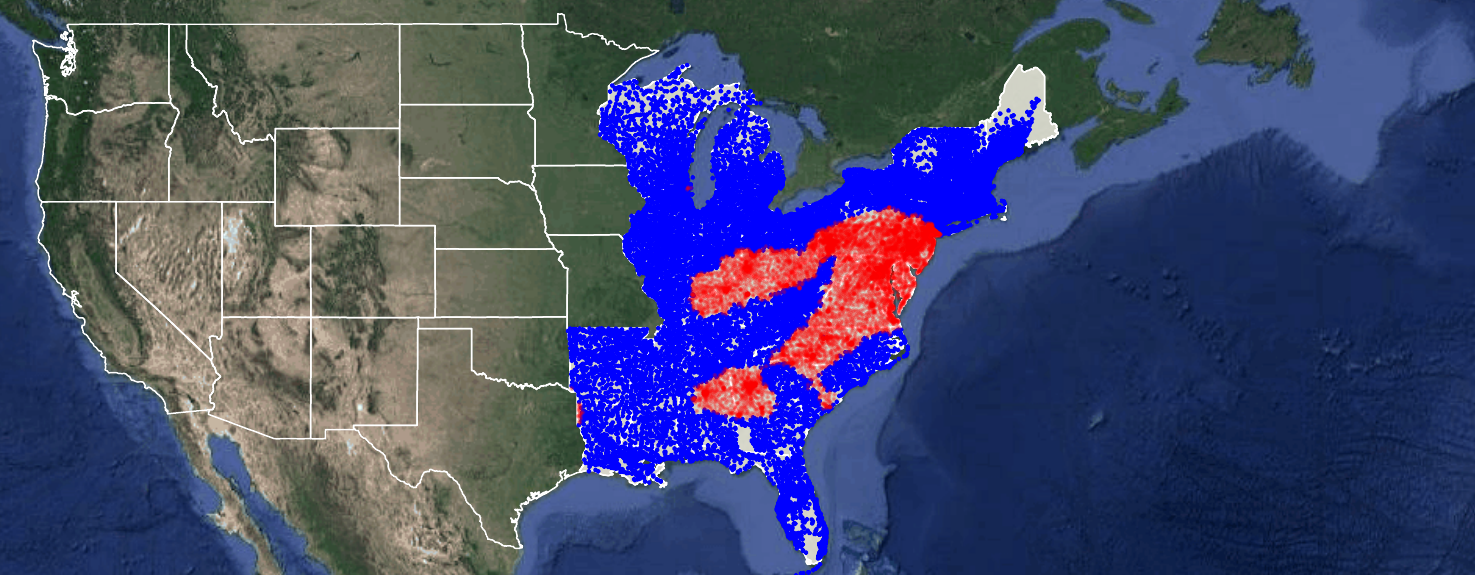}
\end{subfigure}%
~
\begin{subfigure}{0.36\textwidth}
\vspace{0.37in}
\includegraphics{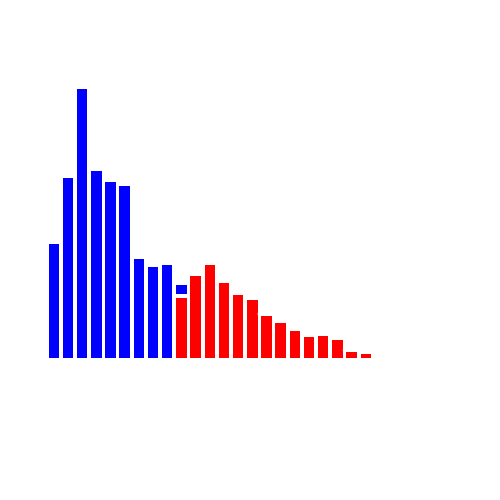}
\end{subfigure}%
\hfill
\color{white}
\begin{subfigure}[t]{0.55\textwidth}
\centering
\includegraphics[trim = {2.1in 3.45in 0cm 4.25in}, clip]{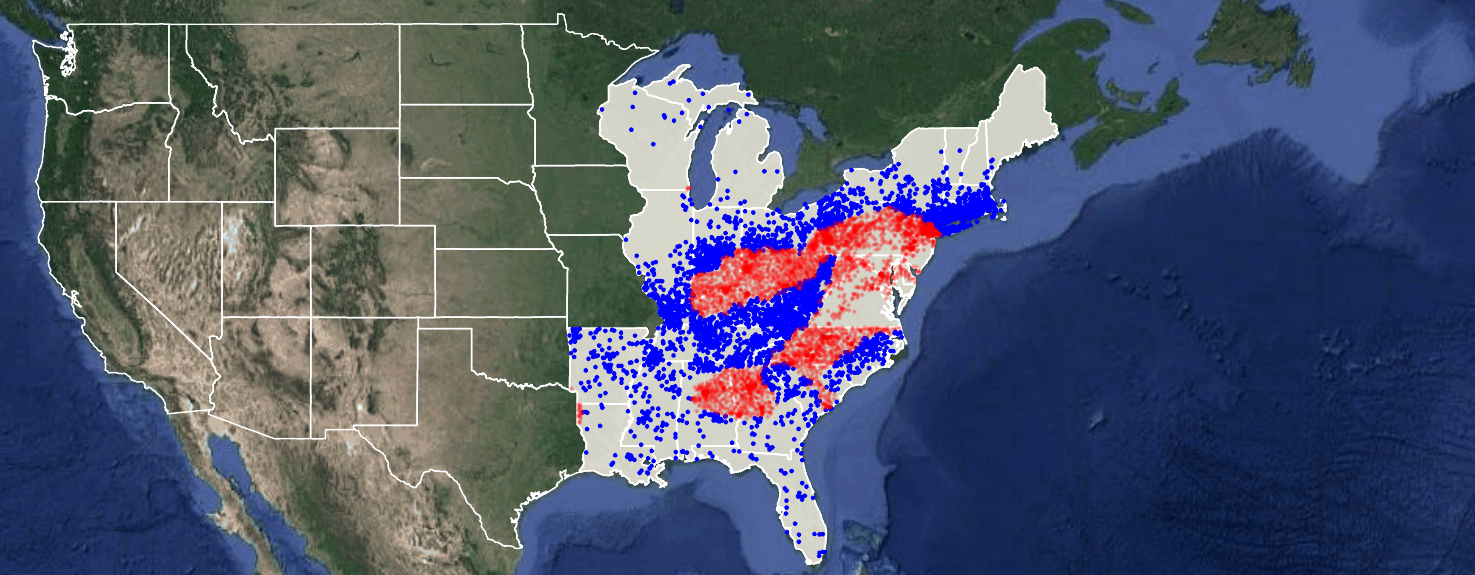}
\end{subfigure}%
~
\begin{subfigure}[t]{0.39 \textwidth}
\vspace{-3cm}
\centering
{\bf
\footnotesize{

\begin{tabular}{@{}lll @{}}\toprule
ZIP Codes & Controls & High \\\midrule
All & 14726 & 6625 \\ 
Matched & 3720 & 3720 \\ 
Unmatched & 8849 & 2325 \\ 
Discarded & 2157 & 580 \\ \bottomrule
\end{tabular}
}}
\end{subfigure}
\color{black}
\caption{Top - InMAP estimates of the influence of emissions from 783 coal-fired generating units (2005) on ZIP codes in the eastern U.S.. Middle - high-exposed (red) and control (blue) locations in the full data set.  Bottom - high-exposed (red) and control (blue) locations in the propensity score matched data set.}
\label{fig:USmap_cont}
\end{figure}

Figure \ref{fig:USmap_cont} (top panel) depicts the InMAP estimates of the influence of coal emissions on each ZIP code. 
We selected a cutoff value of \SI{4.0}{\micro\gram\per\metre^3} in the coal emissions influence to classify locations as either high-exposed or controls because this value is between the two modes in the distribution (Figure \ref{fig:USmap_cont}: top).  Using this cut-off, there were 6,625 high-exposed and 14,726 control locations.  In terms of coal emissions influence, the high-exposed locations were in the 80th percentile of all U.S. ZIP codes and in the 69th percentile of ZIP codes included in the study area.  

\subsection{Propensity Score Matching Results}


High-exposed locations were matched to controls in the same region with similar propensity scores.  Figure \ref{fig:USmap_cont} (bottom) depicts the locations of ZIP codes in the propensity score matched data.  The matched data set consisted of 3,720 high-exposed and 3,720 controls, which is 35\% of the original 21,351 locations. There were 190,339 IHD events in over 6 million person-years in the matched data set.  IHD rates per 100,000 person-years in the high-exposed (controls) were 3170 (3162) in the Industrial Midwest, 2843 (2589) in the Northeast, and 2853 (2720) in the Southeast. Standard diagnostics for propensity score matching were performed (see Appendix \ref{app:matching}) to ensure the comparability of the high-exposed and control locations, thus mitigating the threat of confounding from these variables.

\subsection{Analysis of IHD in the Matched Data}

Using Poisson regression fit to the matched data in each region, the estimated IRRs for annual IHD hospitalizations comparing high-exposed locations to controls were 1.016 (95\% CI: 0.998, 1.035) in the Industrial Midwest, 1.077 (95\% CI: 1.060, 1.094) in the Northeast, and 1.058 (95\% CI: 1.042, 1.075) in the Southeast.  This indicates a significant increase in the rate of IHD hospitalizations in high-exposed locations in the Northeast and Southeast after adjusting for covariates.  There was no significant increase in the Industrial Midwest.  

\subsection{Sensitivity to Propensity Score Procedure}

Two sensitivity analyses, one stratifying on estimated propensity scores instead of matching, and another using DAPS matching, resulted in similar IRR estimates as the primary analysis. The stratified analysis included 18,614 ZIP code locations, 87\% of the full data set, in the IHD model. For further details and results on DAPS matching, see Appendix \ref{app:DAPS}.

\subsection{Sensitivity to the Definition of High vs. Low Exposure}
\begin{figure}
\begin{subfigure}{\textwidth}
\centering
\includegraphics{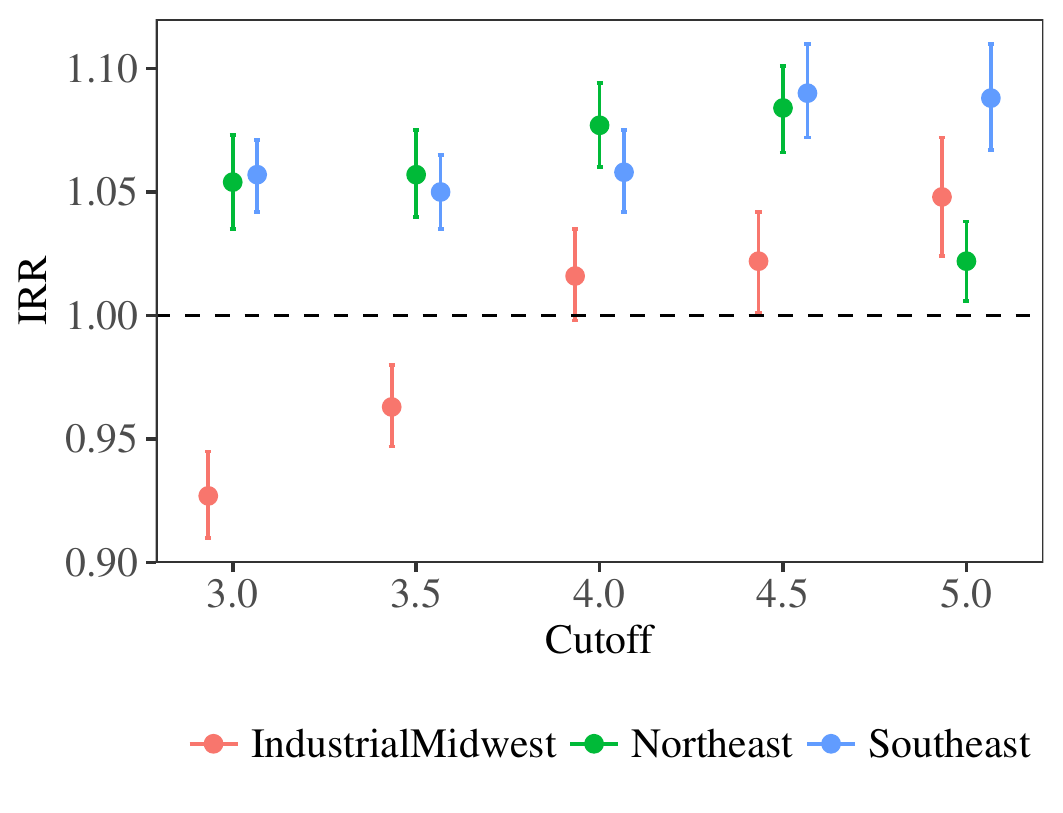}\\
\end{subfigure}
~
\begin{subfigure}{\textwidth}
\centering
\includegraphics{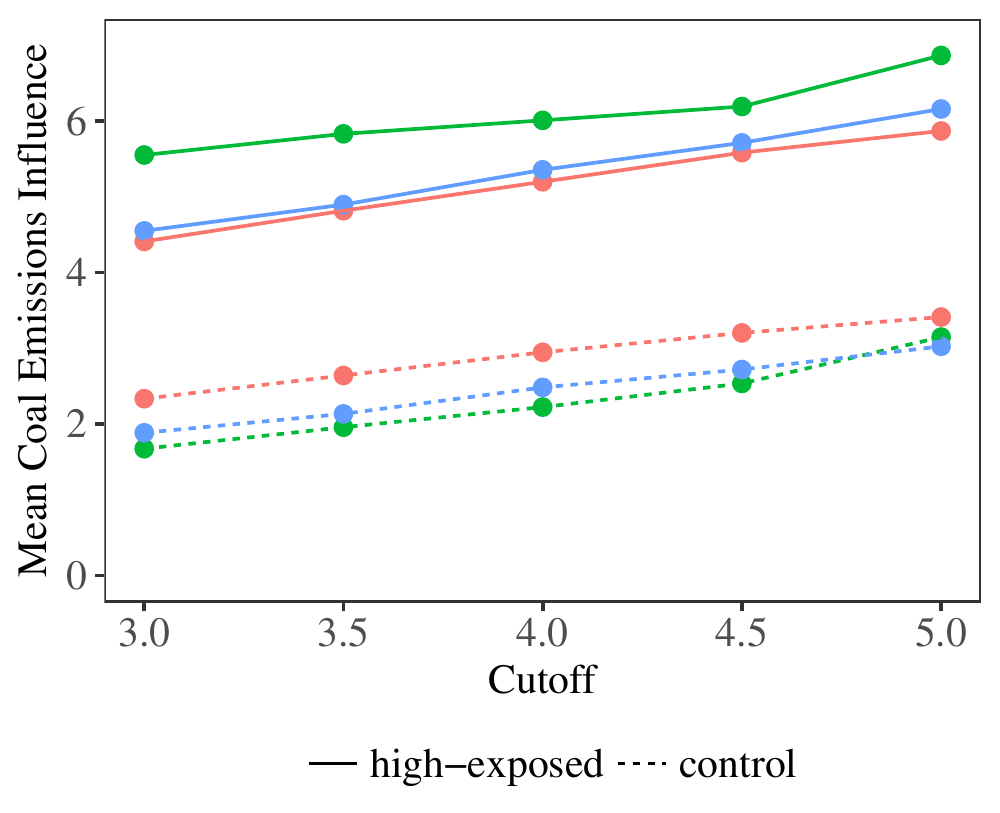}
\end{subfigure}
\caption{Top - estimated IRRs associated with IHD hospitalizations comparing high-exposed locations to controls for various cutoffs. Bottom - mean coal emissions influence by exposure group at the cutoffs.}
\label{fig:effects}
\end{figure}

Figure \ref{fig:effects} (top) shows estimated IRRs for the exposure/IHD relationship for various cutoffs ranging from $3.0 \text{ to } \SI{5.0}{\micro\gram\per\metre^3}$ for delineating high coal emissions exposed ZIP codes from controls.  This range was selected because these cutoff values are between the two modes of the coal emissions influence distribution.  In the Northeast and Southeast, positive associations were observed at every cutoff.  In the Industrial Midwest, lower cutoffs resulted in negative associations.  Larger cutoffs in that region resulted in either small or no effects with the point estimates becoming more positive with higher cutoffs.

Figure \ref{fig:effects} (bottom) shows the mean coal emissions influence in the high-exposed and controls, by region, for the range of cutoffs. It depicts the contrasts, in terms of the continuous coal emissions influence, between the two exposure groups.  The high-exposed locations in the Northeast have the highest coal emissions influence and the contrast between the high-exposed and controls is the largest among the regions.  In the Industrial Midwest, the high-exposed areas have a comparatively low coal emissions influence and the contrasts are the smallest. Note the IRR estimates increase with larger contrasts in the Industrial Midwest.

\subsection{Secondary Analysis Results: Adjusting for \PM{}}

\begin{table}[!h]
\centering
\begin{tabular}{@{}l l l l@{}}\toprule
Analysis & Industrial Midwest & Northeast & Southeast \\\midrule
Primary & 1.02  & 1.08  & 1.06 \\
& (1.00, 1.04) & (1.06, 1.09) & (1.04, 1.08) \\\cmidrule{2-4}
Secondary & 1.01 & 1.02  & 1.05  \\
&(0.99, 1.03) & (1.01, 1.04) & (1.03, 1.07)\\\bottomrule
\end{tabular}
\caption{Estimated IRRs associated with IHD hospitalizations in the primary (\PM{} unadjusted) and secondary (\PM{} adjusted) analyses.}
\label{tab:secondary}
\end{table}

Table \ref{tab:secondary} contains estimated IRRs for IHD hospitalizations comparing high-exposed to control locations in both the primary (\PM{} unadjusted) and secondary (\PM{} adjusted) analyses, showing that the secondary analysis yields attenuated IRRs in each region. In interpreting these results, it is important to consider the relationship between \PM{} and the coal influence exposure in each region.  

In the Southeast, total \PM{} mass concentrations in the propensity score matched data were only slightly higher (see Appendix \ref{app:secondary}) in the high-exposed locations than the controls suggesting that high-exposed and control areas are differentiated by the amount of coal-derived \PM{}, even though total annual average \PM{} mass is similar.  This provides some support for interpreting results from the secondary analysis as effects of evelaved coal-derived emissions in areas with similar total \PM{} mass.  In other words, in the Southeast, elevated coal emissions influence may be associated with increased IHD above and beyond what might be due to associations with overall \PM{} mass, possibly due to other characteristics of the \PM{}. The lack of an association in the Industrial Midwest precludes a similar interpretation of the secondary analysis in that region.


In the Northeast, however, there was a more pronounced association between high-exposure and total \PM{} concentrations (see Appendix \ref{app:secondary}), reflecting the prominent influence of coal emissions in determining overall \PM{} mass in this region.  As a result, the IRR effect estimate in the secondary analysis was particularly attenuated (primary IRR: 1.08, secondary IRR: 1.02). Given the lack of independence between exposure and total \PM{} concentrations, we cannot interpret the secondary analysis as being informative about the relative toxicity of particles from coal emissions.


\section{Discussion}

We have deployed new computational and statistical tools to investigate the health impact of exposure to emissions originating from coal power plants.  For this analysis, we considered populations in 21,351 ZIP code locations in the Northeast, Industrial Midwest, and Southeast regions of the United States and their exposure to the emissions from 783 coal-fired generating units in 2005.  Our results showed an increased rate in IHD hospitalizations in the Northeast (IRR: 1.08, 95\% CI: 1.06, 1.09) and the Southeast (IRR: 1.06, 95\% CI: 1.04, 1.08) for high-exposed locations.  Importantly, this is an impact on annual hospitalization rates, whereas most previous studies of IHD admissions and air pollution have only addressed short-term exposure and daily rates.  No significant association was found in the Industrial Midwest (IRR: 1.02, 95\% CI: 1.00, 1.04). While there are no directly comparable studies, our results are broadly consistent with existing work. The source-apportinment analysis in Thurston et al (2016) estimated a hazard ratio for IHD mortality of 1.05 (95\% CI: 1.02, 1.08) per $\SI{10}{\micro\gram}/\text{m}^3$ increase in coal combustion \PM{}.\cite{thurston2016ischemic} 

The lack of an association in the Industrial Midwest may be the result of the relative spatial homogeneity in coal power plant exposure in that region, reducing the exposure gradient used for comparison and the ability to detect health effects. The Industrial Midwest had the smallest difference in mean coal emissions influence between the high-exposed and control locations.  
In addition, the highest and least exposed ZIP codes were underrepresented there.  In terms of coal emissions influence, only 1.4\% (5.0\%) of Industrial Midwest ZIP codes were among the highest (lowest) 10\% of ZIP code exposure levels in the propensity score matched data (for comparison, the analogous values in the NE were 17\% (16.5\%)). 

At lower cutoffs, negative associations estimated in the Industrial Midwest can be attributed to controls in the matched data being densely located in two areas with high IHD rates derived from other causes.  Specifically, coal mining in southeast Kentucky and southern West Virginia and steel production near Lake Erie result in locally high IHD rates. \cite{hendryx2009mortality,landen2011coal} At higher cutoffs, these areas, which are consistently characterized as control locations, are not included in the matched data due to more suitable controls being located elsewhere. 


Losses in predictive accuracy are a potential issue for reduced-complexity models like InMAP. It is important to reiterate that our strategy does not use outputs from InMAP directly in the health-outcomes analysis, it merely uses the output to characterize ZIP codes as ``high-exposed'' or controls.  Thus, our design provides robustness to losses in predictive accuracy from the use of reduced-complexity models  
as we require only accurate estimates of the relative ranking of coal emissions influence among ZIP codes. 

While the propensity score analysis employed here is designed to provide a more rigorous account of the threat of confounding, it comes with important limitations.  The method relies on a binary exposure classification of an inherently continuous exposure.  Dichotomizing the continuous exposure permits more targeted adjustment for confounding, including empirical assessment of the extent to which covariates are ``balanced'' between high and low exposed groups. 
Nonetheless, dichotomization of a continuous exposure results in a loss of information and the inability to characterize a complete exposure-response relationship. It can also be viewed as a type of classical measurement error.  This limitation was evident in the Industrial Midwest, where exposures between the two groups were closer on a continuous scale than the other regions.  


Another inherent feature of the investigation is the presence of ambient \PM{} mass as an intermediate variable ``on the causal pathway'' between coal emissions influence and IHD hospitalizations.  Our primary analysis did not provide any adjustment for ambient \PM{} mass, permitting IRR estimates to include any effect of high coal emissions influence that is due to the resulting increase in total \PM{} mass.  In a secondary analysis that adjusts for \PM{} in the propensity score and outcome models, we evaluated the possibility of interpreting results as differential effects of coal emissions influence among areas with the same overall \PM{} mass.  We provided some evidence that the assumptions required for such interpretation are reasonable in the Industrial Midwest and Southeast, but more targeted analysis of this point, possibly with methods emanating from the literature on mediation analysis or principal stratification, are warranted. \cite{robins1992identifiability,frangakis2002principal}

\section{Contributors}
CZ developed the analysis plan and provided oversight throughout the study, along with writing large parts of the manuscript and extensive editing.  KC was responsible for data analysis and drafting the first manuscript.  CC provided software to run InMAP directly in the R computing environment and other software tools for assembling the analysis data set.  CK provided code for data linkages and draft editing.  LH provided environmental engineering expertise and contributed significant improvements to initial manuscripts.  JS contributed to a critical review of the analysis and draft editing, as well as presided over material related to the data-fused pollution estimates.  All authors read and approved the final version. 

\section{Declaration of Competing Interests}

We declare no competing interests. 

\section{Acknowledgements}
This publication was made possible by USEPA grant RD-835872-01, NIH grant R01ES026217, and HEI grant 4953.  Its contents are solely the responsibility of the grantee and do not necessarily represent the official views of the USEPA.  Further, USEPA does not endorse the purchase of any commercial products or services mentioned in the publication.  

\noindent We would like to thank Christopher Tessum and Julian Marshall of the University of Washington for their input and insights on the use of InMAP in this study.  KC would like to thank the General Omar N. Bradley Foundation.

\begin{singlespacing}
\bibliographystyle{unsrt}
\bibliography{references}{}
\end{singlespacing}

\newpage
\begin{appendices}
\section{Data}
\label{app:data}

\begin{minipage}{\textwidth}
\centering
\renewcommand{\arraystretch}{0.75}
\begin{tabular}{@{}l l l@{}} \toprule
Abbreviation & Description & Source \\\midrule
\PM{} & Average annual concentration ($\SI{}{\micro\gram}/\text{m}^3$) of \PM{} in 2005 & Di et al\cite{di2016assessing} \\
 & (secondary analysis only) & \\
PctOccupied & Percent of housing units occupied & Census 2000 \\
PctUrban & Percent residing in an urban area & Census 2000\\
logPop & log(total population)  & Census 2000 \\
MedianHHInc & Median household income (thousands of \$) & Census 2000 \\
PctHighSchool & Percent with a high school degree & Census 2000\\
PctFemale & Percent female  & Census 2000 \\
PctBlack & Percent African-American & Census 2000\\
PctPoor & Percent living below poverty threshold & Census 2000\\
PctMovedIn5 & Percent moved in last 5 years & Census 2000\\
MedianHValue & Median house values (thousands of \$) & Census 2000\\
mean\_age & Mean age of the Medicare population & Medicare 2005 \\
Female\_rate & Percent female (Medicare pop.) & Medicare 2005\\
White\_rate & Percent Caucasian (Medicare pop.) & Medicare 2005 \\
avrelh & Average relative humidity (2005) & Di et al\cite{di2016assessing} \\
avtmpf & Average temperature (2005) & Di et al\cite{di2016assessing} \\
smokerate2000 & County smoking rate (2000) & Dwyer-Lindgren \\
& & et al \cite{dwyer2014cigarette} \\\bottomrule
\end{tabular}
\captionof{table}{Covariates included in the propensity score model.}
\label{tab:covariates}
\end{minipage}

\section{Geographic Regions}
\label{regions}

\begin{minipage}{\textwidth}
\centering
\includegraphics[trim = {4.25in 3.55in 1.65in 4.25in}, clip]{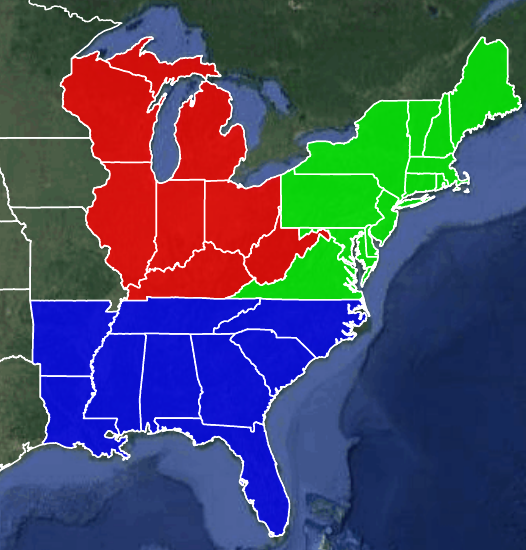}
\captionof{figure}{United States regions (Industrial Midwest, Northeast, Southeast) included in this study.}
\label{regions}
\end{minipage}

\section{Propensity Score Matching}
\label{app:matching}

\begin{figure}
\centering
\includegraphics[scale = 0.80,trim = {0 2cm 0 1cm}, clip]{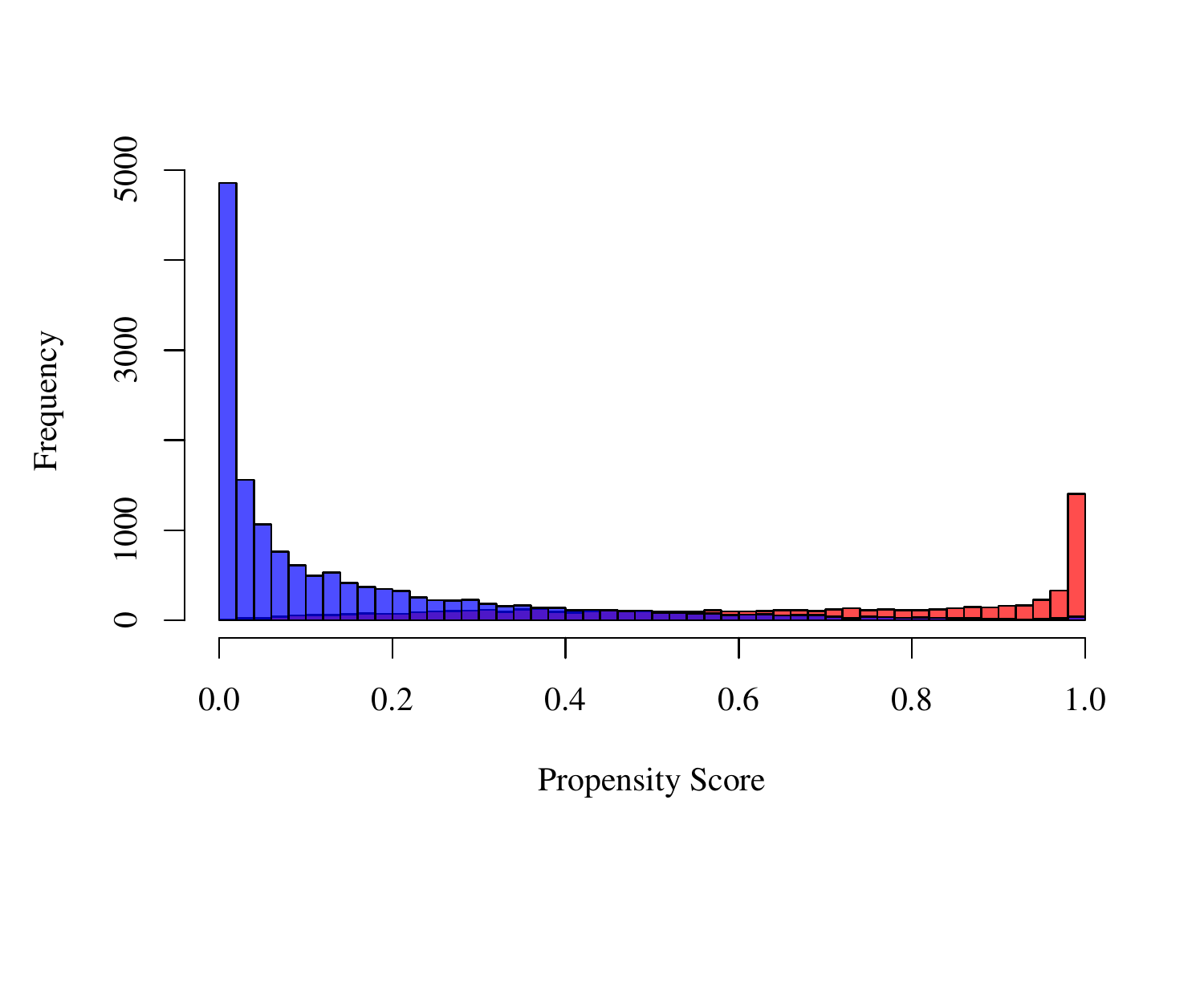}
\caption{Distribution of estimated propensity scores for high-exposed (red) and control (blue) locations before matching.}
\label{fig:PS_dist}
\end{figure}

We estimated the propensity score, which is the predicted probability of being high-exposed conditional on covariates, of each ZIP code using logistic regression.  Figure \ref{fig:PS_dist} shows the distribution of the estimated propensity scores for high-exposed and control locations.  The propensity score distributions are very different, indicating stark differences in the characteristics of high-exposed and control locations and a strong threat of confounding in unadjusted, health-outcome comparisons.    The purpose of the propensity score matching algorithm is to match high-exposed locations to controls with similar propensity scores.  

We used a 1:1 nearest neighbor algorithm with caliper implemented in the R MatchIt package \cite{ho2011matchit}.  The caliper is the maximum allowable difference in propensity scores between matched locations.  We used calipers equal to 20\% of the pooled standard deviation of the logit of the propensity score, as suggested in Austin (2011) \cite{austin2011introduction}.  These calipers were 0.38, 0.61, and 0.31 for the Industrial Midwest, Northeast, and the Southeast, respectively. 

In addition, locations with propensity scores outside the mutual support of the two groups' propensity scores were discarded from the analysis to prevent extrapolation beyond the observed range of covariate profiles common to both exposure groups.

\begin{figure}
\centering
\includegraphics{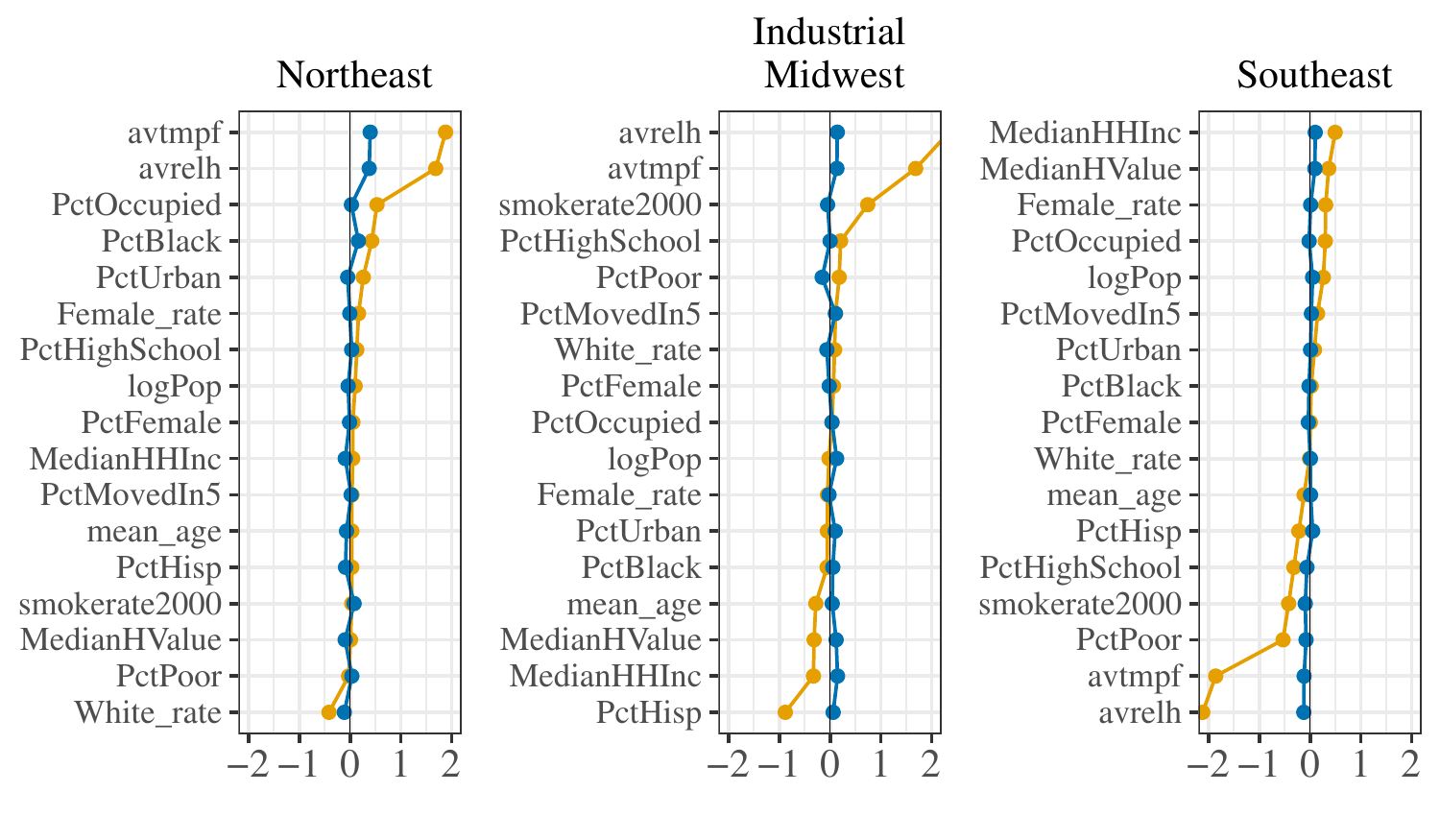}
\caption{Standardized mean difference (SMD) between the high-exposed and control groups for each covariate in the raw (orange) and propensity score matched (blue) data. Variable abbreviations are in Appendix \ref{app:data}.}
\label{fig:SMD}
\end{figure}

The matching process resulted in 3,720 of the 6,625 (56\%) high-exposed locations receiving matches with similar propensity scores. Table \ref{tab:matcheddata} provides descriptive statistics of the matched data set. After matching, we reviewed several diagnostics to ensure the matching process successfully balanced covariates, which would adjust for confounding.  Figure \ref{fig:SMD} depicts one common diagnostic, the standardized mean difference, for each covariate in the raw and propensity score matched data.  The standardized mean difference is the difference in means between the high-exposed and controls, divided by the pooled standard deviation of the two groups \cite{austin2011introduction}.  Differences are close to zero in the matched data for each covariate in each region, except for average temperature and humidity, indicating that covariates in the matched data are ``balanced'' (on average) between high-exposed and matched control locations.   The ability to confirm such balance is a key benefit of using propensity scores.  Average temperature and humidity were adjusted for in the outcome model. Thus, the threat of confounding due to these factors is minimized. 

\afterpage{%
\thispagestyle{empty}
\newgeometry{left = 1in, right = 0.25in, top = 0.25in, bottom = 0.25in}
\begin{landscape}

\begin{minipage}{10.5in}
\makebox[\textwidth][c]{
\renewcommand{\arraystretch}{0.75}
\begin{tabular}{@{} lllllll @{}}
  \toprule
 & \multicolumn{2}{c}{Industrial Midwest} & \multicolumn{2}{c}{Northeast} & \multicolumn{2}{c}{Southeast} \\ 
  \midrule
 &  Controls & High-exposed & Controls & High-exposed & Controls & High-exposed \\\midrule
Number of ZIP codes & 1234 & 1234 & 1266 & 1266 & 1220 & 1220 \\ 
  IHD events  & 19.11 (30.55) & 21.85 (34.96) & 28.95 (35.81) & 29.93 (45.30) & 26.63 (31.09) & 26.86 (32.13) \\ 
  Person-years  & 604.20 (995.62) & 689.26 (1080.49) & 1118.45 (1343.46) & 1052.92 (1521.75) & 979.04 (1143.19) & 941.33 (1100.74) \\ 
  \PM{}  & 14.71 (1.71) & 15.06 (1.36) & 12.38 (1.51) & 13.89 (1.62) & 14.13 (1.49) & 14.53 (1.58) \\ 
  logPop  & 7.77 (1.62) & 7.98 (1.62) & 8.63 (1.53) & 8.56 (1.76) & 8.81 (1.46) & 8.88 (1.39) \\ 
  PctUrban  & 0.29 (0.38) & 0.33 (0.40) & 0.61 (0.43) & 0.59 (0.43) & 0.42 (0.41) & 0.42 (0.41) \\ 
  PctBlack  & 0.03 (0.09) & 0.04 (0.11) & 0.06 (0.13) & 0.09 (0.18) & 0.23 (0.25) & 0.22 (0.23) \\ 
  PctHisp  & 0.01 (0.01) & 0.01 (0.02) & 0.06 (0.11) & 0.05 (0.11) & 0.03 (0.06) & 0.03 (0.04) \\ 
  PctHighSchool  & 0.41 (0.10) & 0.41 (0.10) & 0.34 (0.11) & 0.35 (0.13) & 0.32 (0.09) & 0.32 (0.08) \\ 
  MedianHHInc  & 34.16 (14.06) & 35.86 (11.34) & 50.16 (22.76) & 48.22 (23.25) & 37.04 (13.62) & 38.38 (11.68) \\ 
  PctPoor  & 0.15 (0.10) & 0.14 (0.09) & 0.10 (0.08) & 0.10 (0.09) & 0.14 (0.08) & 0.13 (0.08) \\ 
  PctFemale  & 0.51 (0.04) & 0.51 (0.04) & 0.51 (0.03) & 0.51 (0.04) & 0.51 (0.04) & 0.51 (0.04) \\ 
  PctOccupied  & 0.88 (0.10) & 0.89 (0.10) & 0.89 (0.13) & 0.89 (0.13) & 0.89 (0.08) & 0.89 (0.09) \\ 
  PctMovedIn5  & 0.40 (0.11) & 0.41 (0.11) & 0.38 (0.10) & 0.38 (0.13) & 0.46 (0.10) & 0.46 (0.12) \\ 
  MedianHValue  & 76.54 (53.65) & 80.53 (32.75) & 165.89 (12656) & 155.30 (15.15) & 92.13 (53.69) & 97.46 (43.70) \\ 
  smokerate2000  & 0.30 (0.03) & 0.30 (0.03) & 0.25 (0.03) & 0.25 (0.03) & 0.27 (0.03) & 0.27 (0.03) \\ 
  avtmpf  & 285.20 (1.64) & 285.35 (1.08) & 283.17 (1.46) & 283.84 (1.43) & 289.68 (2.39) & 289.48 (1.62) \\ 
  avrelh  & 0.01 (0.00) & 0.01 (0.00) & 0.01 (0.00) & 0.01 (0.00) & 0.01 (0.00) & 0.01 (0.00) \\ 
  mean\_age  & 74.48 (1.28) & 74.54 (1.35) & 75.36 (1.34) & 75.25 (1.47) & 74.57 (1.16) & 74.58 (1.13) \\ 
  Female\_rate  & 0.55 (0.05) & 0.55 (0.06) & 0.57 (0.06) & 0.57 (0.06) & 0.58 (0.05) & 0.58 (0.05) \\ 
  White\_rate  & 0.96 (0.09) & 0.96 (0.11) & 0.91 (0.18) & 0.88 (0.20) & 0.80 (0.22) & 0.80 (0.21) \\ 
   \bottomrule
\end{tabular}
}
\captionof{table}{Mean (standard deviation) of ZIP code-level variables in the propensity score matched dataset.}
\label{tab:matcheddata}
\end{minipage}
\end{landscape}
\clearpage
\aftergroup\restoregeometry%
}

\section{Distance Adjusted Propensity Score Matching}
\label{app:DAPS}

DAPSm allows the investigator to modify the relative importance of propensity score similarity and geographic distance in selecting matches by specifying a weight between zero (geographic distance matching) and one (propensity score matching). Using DAPSm, instead of propensity score matching, typically results in matched data sets with geographically closer matches, but with some additional covariate imbalance.  In our analysis, we created DAPS matched data sets for a range of weights using the DAPSm package in R \cite{doi:10.1093/biostatistics/kxx074} and selected the largest weight for which the standardized mean difference (SMD) of all covariates was less than 0.15.    

\begin{figure}[h!]
	\color{white}
	\centering
	\begin{subfigure}[t]{0.55\textwidth}
		\centering
		\includegraphics[scale = 1, trim = {2.1in 3.45in 0cm 4.25in}, clip]{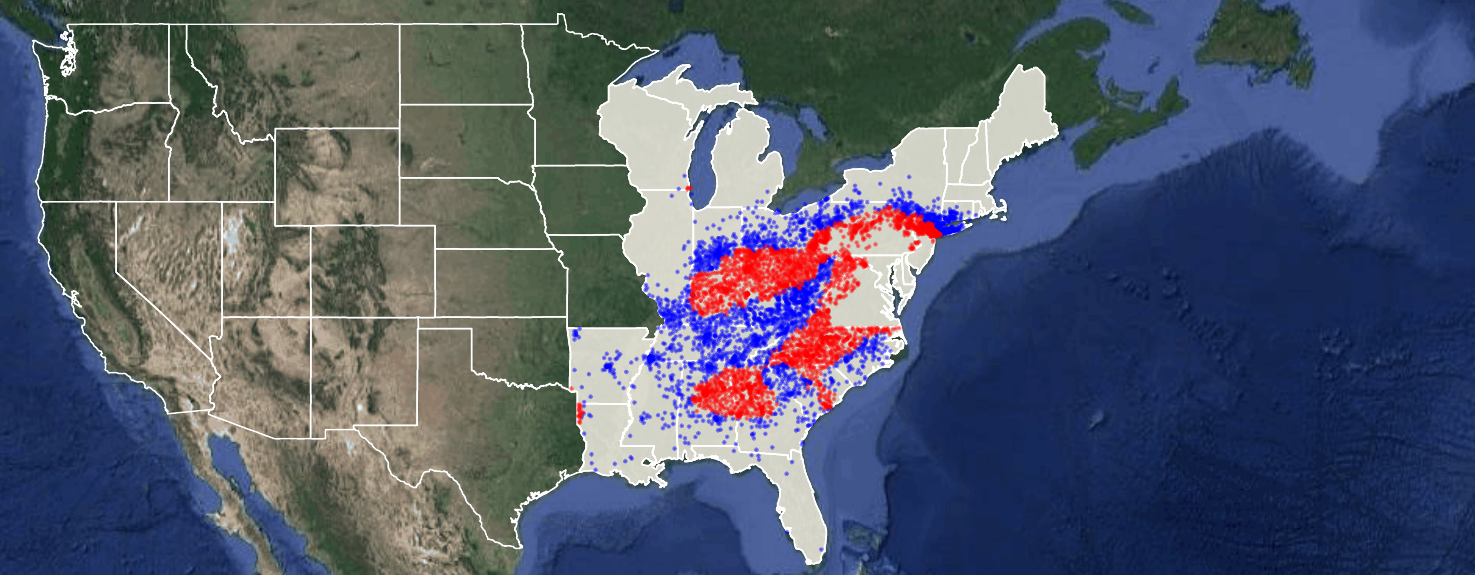}
	\end{subfigure}%
    ~
    \begin{subfigure}[t]{0.39 \textwidth}
    \vspace{-3cm}
    \centering
    {\bf
	\footnotesize{
	
    \begin{tabular}{@{}lll @{}}\toprule
 		ZIP Codes & Controls & High \\\midrule
 All & 14726 & 6625 \\ 
  Matched & 3150 & 3150 \\ 
  Unmatched & 9419 & 2895 \\ 
  Discarded & 2157 & 580 \\ \bottomrule
	\end{tabular}
	}}
	\end{subfigure}
    \color{black}
\caption{High-exposed (red) and control (blue) locations in the Distance Adjusted Propensity Score matched data. 
}
\label{fig:DAPSmap}
\end{figure}

\begin{figure}
\begin{subfigure}{\textwidth}
\centering
\includegraphics[scale = 0.75]{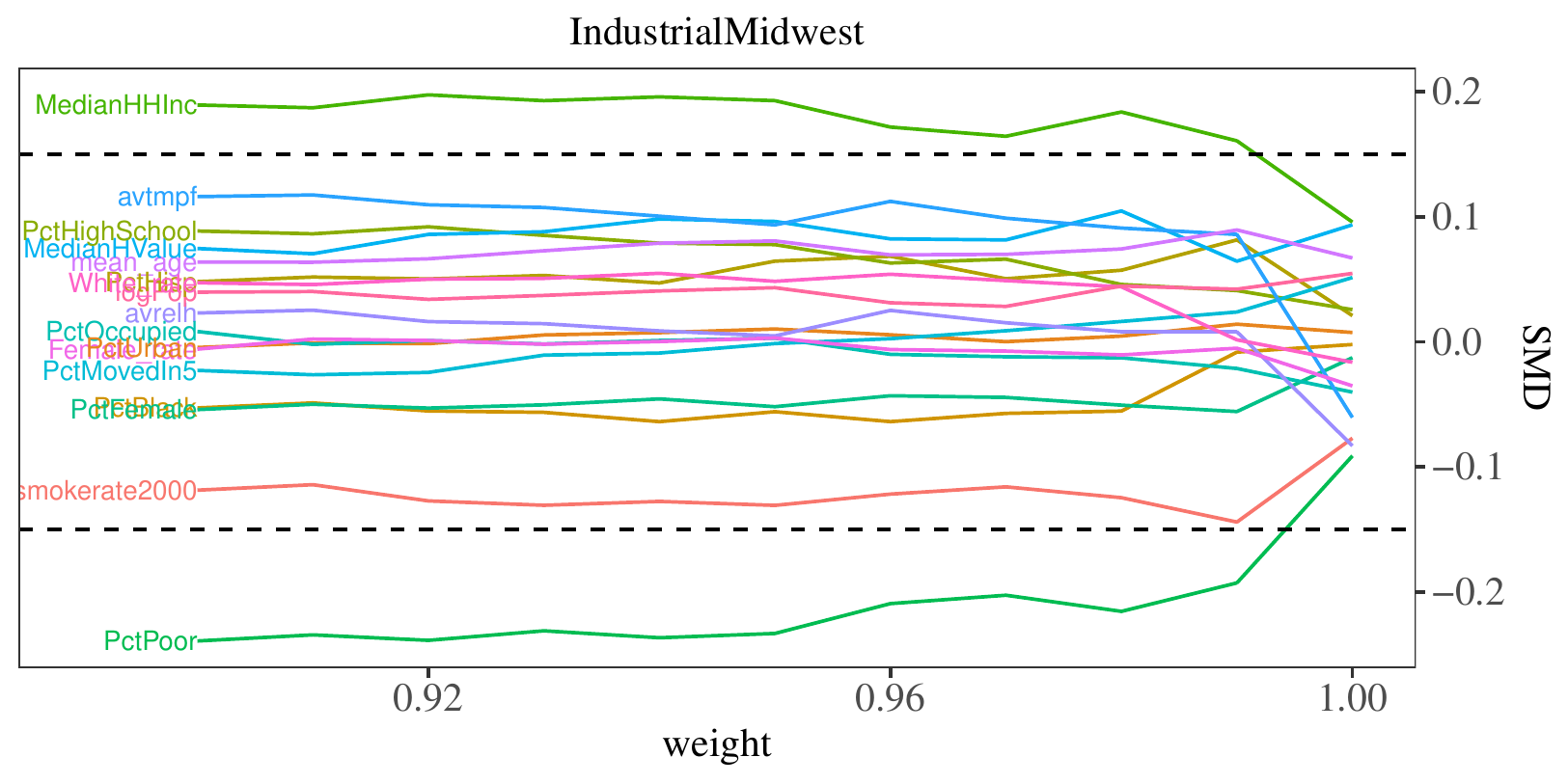}
\end{subfigure}
\begin{subfigure}{\textwidth}
\centering
\includegraphics[scale = 0.75]{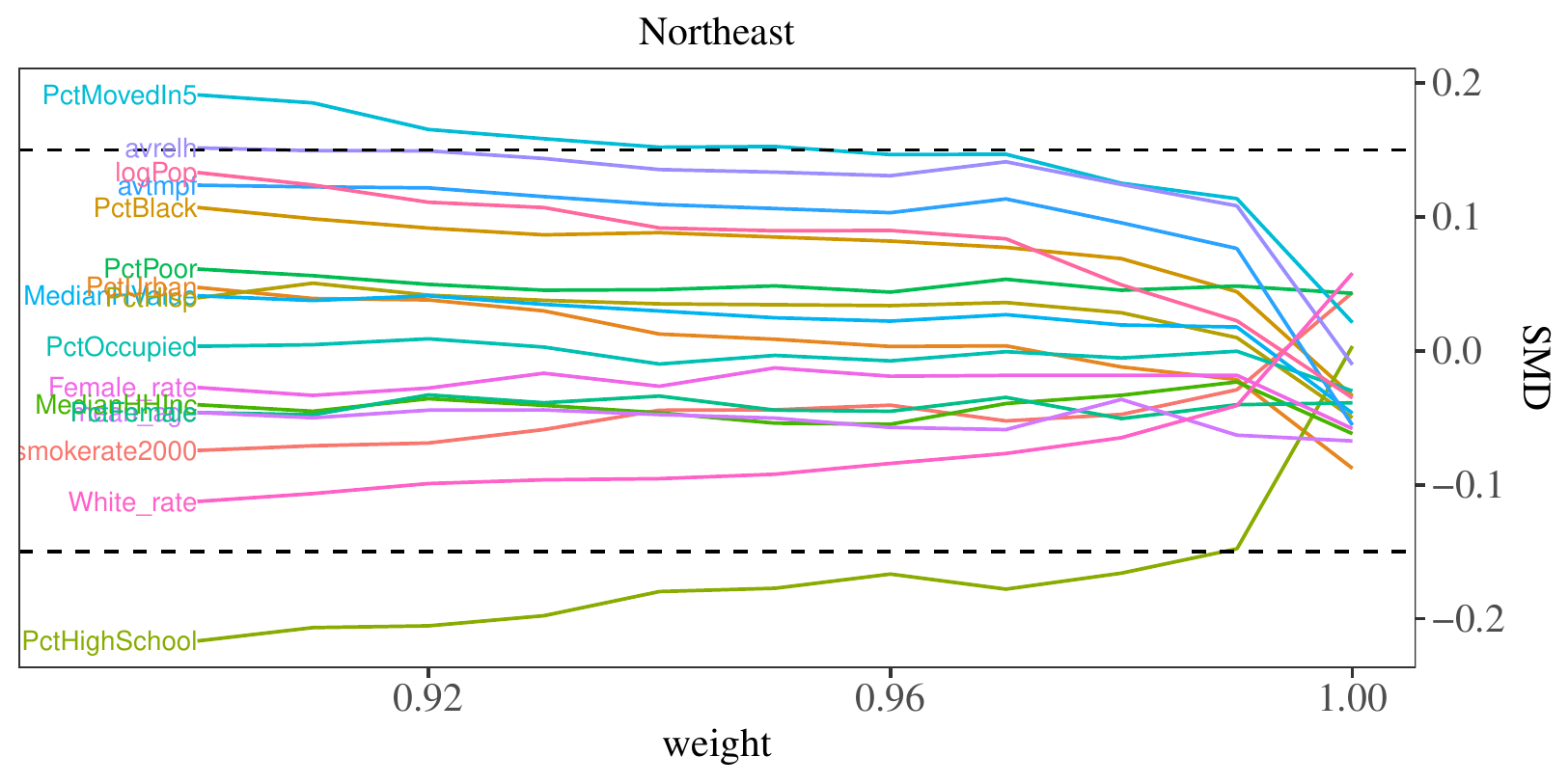}
\end{subfigure}
\begin{subfigure}{\textwidth}
\centering
\includegraphics[scale = 0.75]{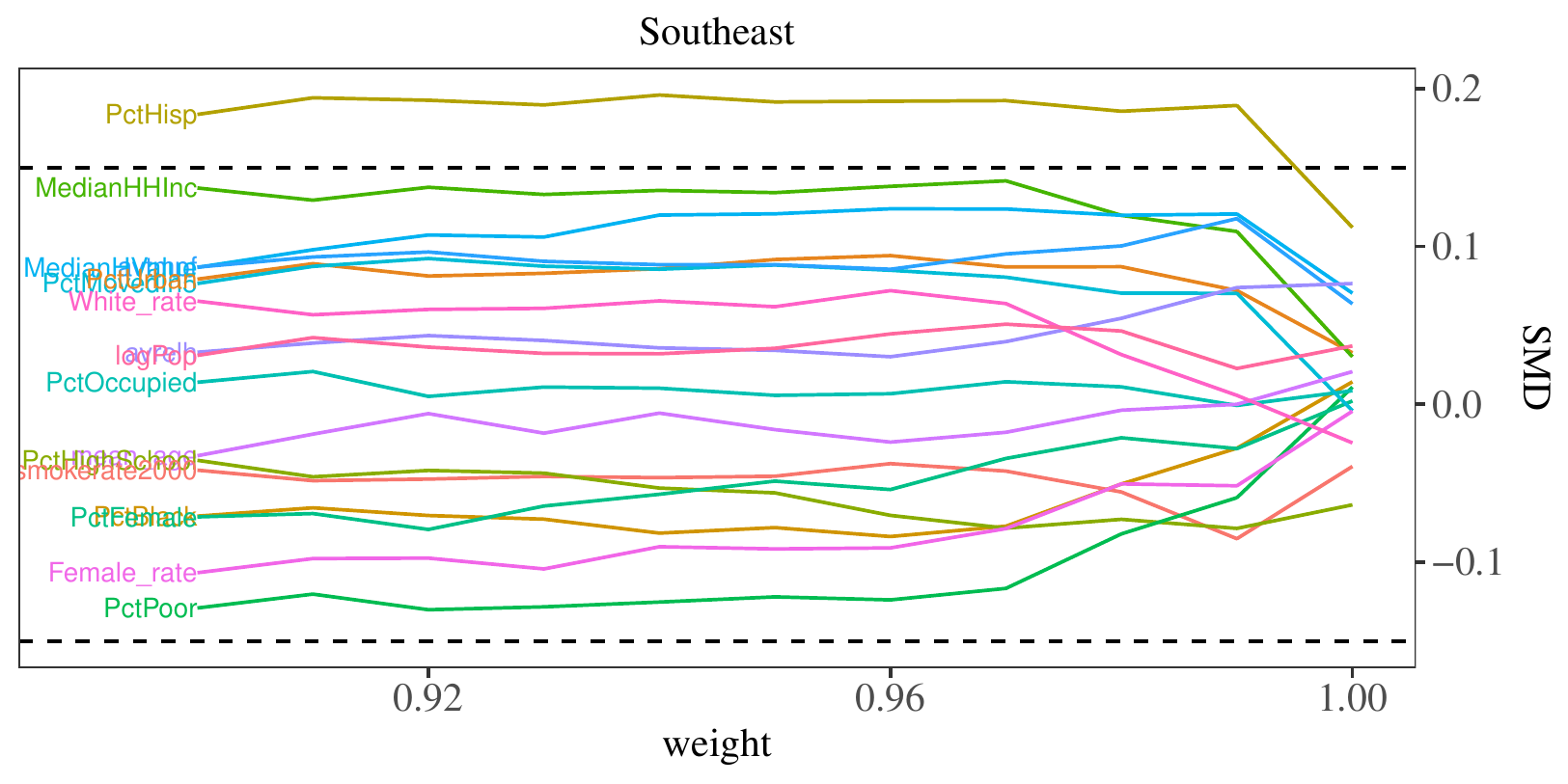}
\end{subfigure}
\caption{Covariate standardized mean differences in DAPSm data sets for various weights.}
\label{fig:DAPS_SMD}
\end{figure}

Figure \ref{fig:DAPS_SMD} shows the SMD of each covariate in the DAPSm data set for a range of weights.  For this sensitivity analysis, we used data sets obtained for weights 0.9975, 0.985, and 0.9975 in the Industrial Midwest, Northeast, and Southeast, respectively. Figure \ref{fig:DAPSmap} depicts the locations of the DAPSm data, which are geographically closer than the propensity score matched data in Figure \ref{fig:USmap_cont}.  Table \ref{tab:DAPSresults} compares the IRRs for IHD estimated using the two matching methods.  

\begin{table}[h!]
\centering
\begin{tabular}{@{}l l l l@{}}\toprule
Analysis & Industrial Midwest & Northeast & Southeast \\\midrule
Propensity Score Matched & 1.02  & 1.08  & 1.06 \\
& (1.00, 1.04) & (1.06, 1.09) & (1.04, 1.08) \\\cmidrule{2-4}
DAPS Matched & 1.00 & 1.06  & 1.05 \\
& (0.98, 1.02)  & (1.04, 1.08)  & (1.03, 1.05) \\\bottomrule
\end{tabular}
\captionof{table}{Comparison of the propensity score and DAPS matched estimates of IRRs for IHD hospitalizations associated with high-exposure to coal power plant emissions.}
\label{tab:DAPSresults}
\end{table}

\section{Secondary Analysis}
\label{app:secondary}

In interpreting the secondary analysis, it is important to consider the relationship between the exposure and total \PM{} mass concentration.  When total \PM{} mass concentrations are similar in the high-exposed and controls, the secondary analysis can be interpreted as the effects of coal power plant influence among areas with similar total \PM{} mass, indicating characteristics of the coal-derived \PM{} itself, other than just total mass, may be responsible for increased IHD. 

To assess this relationship, a common measure is the standardized mean difference, which is the difference in means between the high-exposed and controls, divided by the pooled standard deviation of the two groups \cite{austin2011introduction}.  The standardized mean differences comparing total \PM{} mass concentration in the high-exposed and controls were 0.23, 0.96, and 0.26 in the Industrial Midwest, Northeast, and Southeast, respectively.  Further interpretations of these results are provided in the main text.
\end{appendices}

\end{document}